\documentclass[twocolumn]{aastex63}

\received{-}
\revised{-}
\accepted{-}

\submitjournal{ApJL}

\shorttitle{Observational Nonstationarity of AGN variability}

\shortauthors{Caplar et al.}

\graphicspath{{./}{figures/}}

\begin{document}

\title{Observational Nonstationarity of AGN Variability: The Only Way to Go Is Down!}
% \title{Long term AGN variability : The only way is down!}

\correspondingauthor{Neven Caplar}
\email{ncaplar@princeton.edu}

\author[0000-0003-3287-5250]{Neven Caplar}
\affiliation{Department of Astrophysical Sciences, Princeton University, 4 Ivy Ln.,  Princeton, NJ 08544, USA }

\author[0000-0002-0033-5041]{Theodore Pena}
\affiliation{Department of Physics and Astronomy, Tufts University, Medford, MA
02155, USA}

\author[0000-0001-9487-8583]{Sean D. Johnson}
\affiliation{Department of Astrophysical Sciences, Princeton University, 4 Ivy Ln.,  Princeton, NJ 08544, USA }

\author{Jenny E. Greene}
\affiliation{Department of Astrophysical Sciences, Princeton University, 4 Ivy Ln.,  Princeton, NJ 08544, USA }

\begin{abstract}

To gain insights into long-term variability of Active Galactic Nuclei (AGN), we analyze an AGN sample from the Sloan Digital Sky Survey (SDSS) and compare their photometry with observations from the Hyper Suprime-Cam survey (HSC) observed $\langle 14.85 \rangle$ years after SDSS. On average, the AGN are fainter in HSC than SDSS. We demonstrate that the difference is not due to subtle differences in the SDSS versus HSC filters or photometry. The decrease in mean brightness is redshift dependent, consistent with expectations for a change that is a function of the rest-frame time separation between observations. At a given redshift, the mean decrease in brightness is stronger for more luminous AGN and for objects with longer time separation between measurements.  %The mean decrease in brightness with time \textbf{indicates that AGN variability is stationary on the timescales longer than the timescales of our observations.}violates the stationarity assumption often invoked in AGN variability studies. 
We demonstrate that the dependence on redshift and luminosity of measured mean brightness decrease is consistent with simple models of Eddington ratio variability in AGN on long (Myr, Gyr) timescales. We show how our results can be used to constrain the variability and demographic properties of AGN populations.
\end{abstract}


\section{Introduction}

Changing flux levels with time are nearly ubiquitous among Active Galactic Nuclei (AGN). Studies of these luminosity fluctuations, i.e., AGN variability, have enabled measurements of central supermassive black hole masses (e.g., \citealp{Ben15}), added insights on the structure of AGN accretion disks (e.g., \citealp{Fau16}), and provided powerful AGN selection techniques (e.g., \citealp{
Schmi10}). AGN variability has been directly observed in large samples on timescales ranging from minutes to days, years, and decades 
(e.g., \citealp{ Mac10, Mac12, Mor14, Car15, Cap17, Smi18}). Through indirect methods and simulations, AGN variability has also been studied on Myr and Gyr scales (e.g., \citealp{Nov11, Bla13, Sar18}). \par
Most of the direct observational studies mentioned above quantify AGN variability as a weakly stationary process, i.e.,  with the assumption that the mean luminosity of statistically large ensembles of AGN does not change with time. Empirically, stochastic variability measured in these short-term studies dominated any possible subtle changes in the mean brightness occurring during the duration of the studies. From theoretical grounds, the stochastic variability is thought to be reflective of the details of the physics of AGN accretion disks and other nearby structures, while the mean change of luminosity would be connected to long timescale accretion processes thought to have minimal impact on typical survey timescales (but see \citealp{Law18}). \par

The assumption of no change of mean brightness on short timescales differs from the long-term studies of AGN activity, which indicate large changes in AGN activity on Myr and Gyr scales. The firmest observational proof comes from the studies of individual extended AGN photoionized clouds, so-called  ``Voorwerp'' objects (e.g., \citealp{ Sar16, Johnson:2018}), and the He\,II transverse proximity effect \citep[e.g.,][]{Schmidt:2018}, which clearly show that some AGN exhibit order-of-magnitude changes in their luminosity on 10$^4$-10$^5$ yr time-scales. \par %Further insights into long-term AGN variability and hence black hole accretion modes require statistical samples of representative AGN. %The support for large amounts of variability on long timescales also comes from simulatory and modelling work (e.g., \citealp{Nov11, Hic14}). \par

There has been comparatively little observational research on the deviations from the symmetric behavior of AGN variability. Numerous early studies conducted observationally difficult searches for potential differences in the  variability properties of AGN that were becoming brighter or dimmer, but found little or no evidence for statistical differences in the properties of AGN light curves that were fading or getting brighter \citep{deVri03, deVri05, Bau09, Voe11}. \cite{Mac12} combined the earliest statistically significant sample of AGN measurements from the Palomar Observatory Sky Surveys (POSS) with the SDSS data. They noted that objects from POSS are dimmer when observed in SDSS. They concluded that this may be explained by a ``Malmquist-like'' bias, i.e., the fact that luminosity-selected sample of variable objects will necessarily be dimmer in the later survey, even if there is no change in the mean brightness of the underlying sample. A similar conclusion was reached by \cite{Rum18} who studied examples of extreme variability by comparing SDSS and Dark Energy Survey measurements. \cite{Mor14} also found the decrease of the mean brightness on decade timescales, for the sample of AGN from the SDSS observed in Pan-STARRS1, but attributed this effect to the filter differences. 

\begin{figure*}
    \centering
    \includegraphics[width=\linewidth]{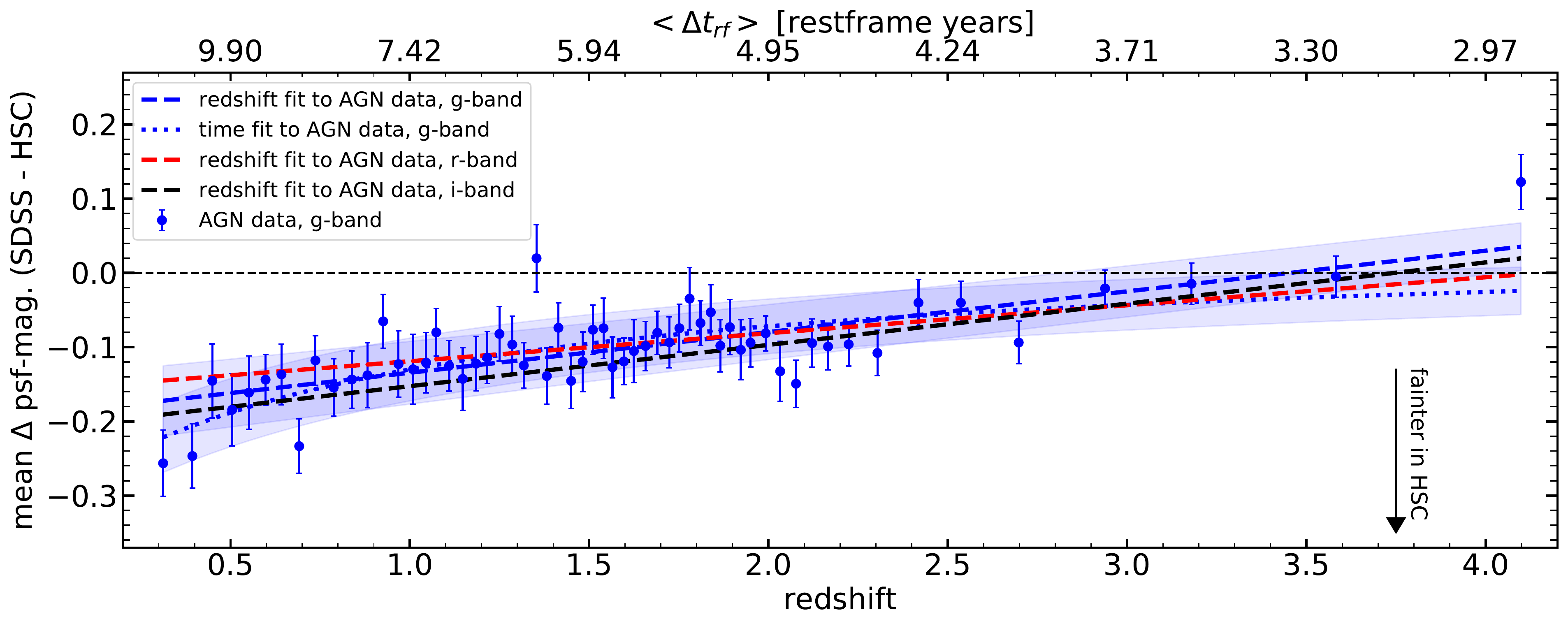}
    \caption{Mean difference in the measured psf-magnitudes for the sample of AGN from SDSS that have been observed in both SDSS and HSC. Blue points show the data for the g-band, while the blue dashed line and the shaded region show the linear fit as a function of redshift to the data and 1-$\sigma$ uncertainty  band. The red and the black dashed lines show the linear fit as a function of redshift in r- and i- bands, respectively. We do not show the data and uncertainty bands for r- and i-band to improve the clarity of the figure, but these are comparable to the g-band quantities. The dotted line shows a fit to the g-band data as a function of restframe time-separation between measurements, indicated on the upper axis. } 
    \label{fig:master}
\end{figure*}

 Here, we use the AGN sample from SDSS and measure their mean brightness in SDSS and HSC. The depth, size, and time separation from SDSS, and the quality of the HSC survey make it especially suitable for this kind of study. In this work, we aim to show that AGN exhibit changes in their mean brightness in a redshift and luminosity dependent manner on the timescales accessible with past (SDSS) and current (HSC) surveys. \par

The code and the data needed to reproduce all of the results mentioned in this work are available at \href{https://github.com/nevencaplar/AGN-Going-Down}{github.com/nevencaplar/AGN-Going-Down}.

\section{Observations}
\label{sec:Observations}

\subsection{Data}

To study AGN variability on decade timescales, we identified AGN from the SDSS \citep[][]{York:2000} DR7 Quasar catalog \citep{Schne10} that were also observed later by the Hyper Suprime-Cam \citep[HSC;][]{Miy18} Subaru Strategic imaging survey. The SDSS survey used a dedicated 2.5 m \citep{Gun10} telescope at Apache Point Observatory to obtain images in five optical bands (ugriz) over a large patch ($\sim$10000 deg$^{2}$) of the northern sky. For a presentation of the photometric calibration and selection function of objects, we refer the reader to the detailed discussion in \cite{Schne10}. The HSC survey is a wide-field optical imaging program being conducted with the 8.2 m Subaru telescope. The second public data release \citep[made available in May 2019;][]{Aih19} covers around 300 deg$^{2}$ overlapping with the SDSS footprint. The HSC data in five optical bands (grizy) are sensitive down to  $\approx26$th magnitude. \par

We searched for objects from the SDSS AGN catalog in the HSC data and recorded their $g$, $r$ and $i$ psf-magnitudes. We excluded all of the objects with any flags showing problems in the calibration. This conservative cut ensures that our conclusions are not driven by possible problems in the brightness measurements in the HSC pipeline. This procedure yields 5919 matched AGN found in both surveys. \par

\subsection{Main result}
To measure the mean difference in the brightness between the two surveys, we split the sample in bins of redshift, each consisting of 100 objects. This number enables us to follow the redshift evolution of the trends in some detail, while minimizing the statistical uncertainty in the mean brightness change. For each redshift bin, we then measured the mean and the median difference between the observed psf-magnitudes in the SDSS and HSC surveys. We also verified that our conclusions are unchanged when using fixed 3 arcsec aperture magnitudes.\par

The resulting mean change in flux is plotted as a function of redshift in Figure \ref{fig:master}. To avoid cluttering the plot, we only show data points at each redshift for measurements in the g-band.  We estimated uncertainties on the mean value at each redshift by bootstrapping the underlying 100 AGN in each bin. We choose to present g-band variability given that contribution of the host-galaxy light, however small for these bright AGN, will be smallest in the bluest available band. However, results for all three bands are very similar. We also show linear fits to the data in all three bands, where one can explicitly see the similarity between all of the results. \par 
We have also verified that the redshift evolution effect is present if we use median differences instead of mean differences of magnitudes, but the magnitude of the effect is somewhat decreased. For instance, the best fit for the median difference is $-0.139+0.051 z$, while for the mean difference it is  $-0.176+0.06 z$. The fact that the effect is still present when using the median shows that it cannot be fully explained by a relatively small number of extremely variable quasars (e.g., \citealp{Mac16,Rum18}). We also show a linear fit to the data as a function of rest-frame time separation between two measurements, i.e., as a function of $14.85 \mbox{ years}/(1+z)$, where 14.85 years is the mean time separation between observations (see Section \ref{sec:time}). This fit also provides a good explanation for the observed data. We discuss the proposed model in which measured changes of the mean/median flux are the consequence of the long-term AGN behaviour and primarily depend on the rest-frame time separation between the two measurements  further in Sections \ref{sec:bri}, \ref{sec:time} and \ref{sec:Modelling}. \par 

To ensure that the observed redshift dependence is not a spurious artifact due to differences between the two surveys, we conduct four different checks that we list here:
\begin{itemize}
    \itemsep0em
    \item consideration of filter differences;
    \item constructing a control sample;
    \item separating the AGN sample according to brightness; and
    \item separating the AGN sample according to the time separation between the SDSS and HSC observations.
\end{itemize}

We elaborate on each of these procedures in some detail below. For consistency, we always show the mean difference in the g-band and conduct linear fits as a function of redshift, but all of our conclusions are applicable to all three bands and fitting variables (redshift or rest-frame time separation)\footnote{Figures showing results for all of the possible combinations of choices for the used observed bands, fitting variables, and using mean/median to derive results can be created from the code and the data available in the GitHub repository}.

\subsection{Filter difference and control sample} 

\begin{figure*}
    \centering
    \includegraphics[width=\textwidth]{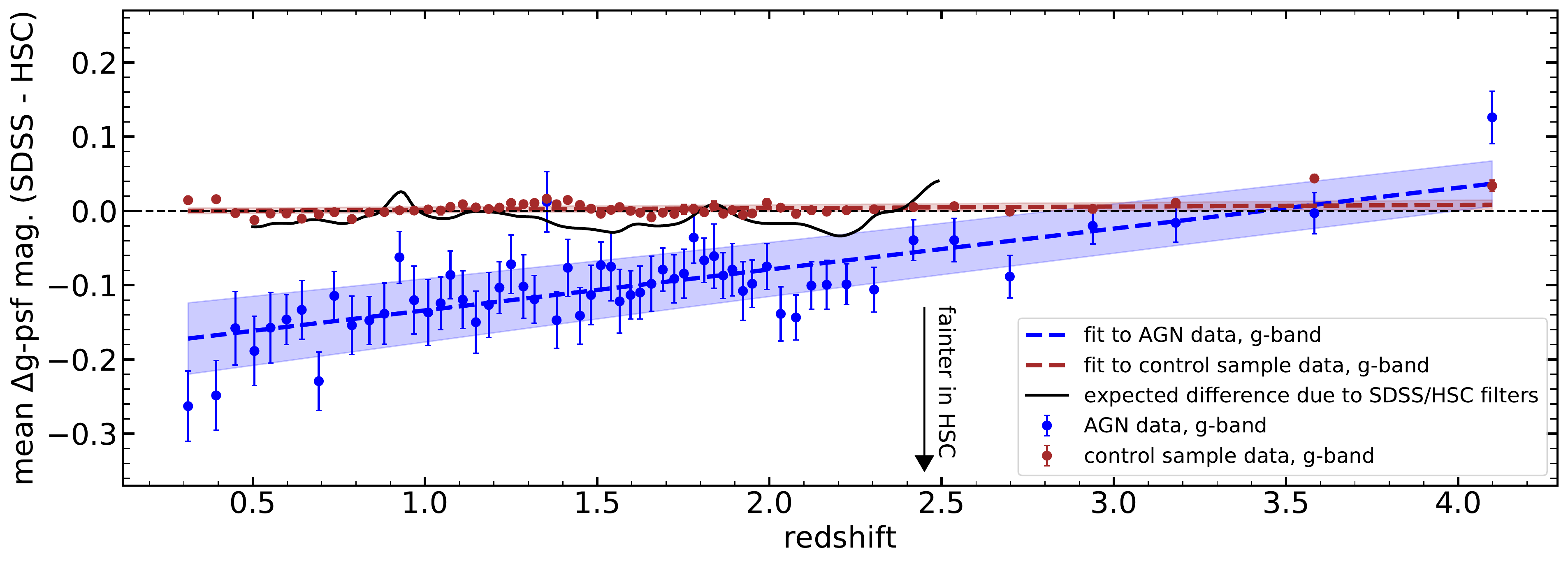}
    \caption{Mean difference in the measured psf-magnitudes for the sample of AGN from SDSS, and the control sample of stars that match these AGN in color. The blue points show the data for AGN, while the blue line and the shaded region show the linear fit to the data and 1-$\sigma$ uncertainty. This is equivalent to the data and fit shown in Figure \ref{fig:master}. The maroon points, line, and shaded region show equivalent quantities constructed for the sample of nonvariable stars in Stripe 82 region. The black line shows the expected redshift dependence due to filter differences between the two surveys. }
    \label{fig:master_filter_stars}
\end{figure*}

We performed two experiments to assess the potential impact of differences in the photometry between the SDSS and HSC surveys that could lead to spurious apparent change of measured brightness. First, to assess the potential impact of differences in the filter systems on the measured magnitudes, we predicted $g_{\rm SDSS} - g_{\rm HSC}$ for the mean SDSS quasar spectral energy distribution \citep[][]{Vanden-Berk:2001} as a function of redshift using the SDSS \citep[][]{Fukugita:1996} and HSC \citep[][]{Kawanomoto:2018} defined system throughput including the filters, telescopes, cameras, and the survey standard atmospheres. Second, we constructed a control sample consisting of nonvariable stars with colors similar to the AGN. The stars were taken from the catalog of nonvariable objects from the equatorial Stripe 82 presented in \cite{Ive07} which we additionally cleaned by removing suspected AGN from \cite{Fle15}. For each AGN we find the star (repetition allowed) that minimizes the Euclidean distance between the measured magnitudes in the g-, r- and i- bands from SDSS. After that, we treated the resulting catalog of stars in exactly the same way as we have treated the AGN sample. As, by definition, we expect no change in the brightness of these stars when imaged in the two surveys, any systematic differences between the two surveys will be expressed in this comparison. These experiments capture effects both from filter differences and from any differences in the PSF magnitude measurement techniques. \par

We show the results of this experiment and deduced effects of filter differences in Figure \ref{fig:master_filter_stars}. The overall decrease in mean flux is not present in the control sample of nonvariable stars. In particular, we emphasize the absence of ``redshift'' trend in the control sample. This is an expected result, as the different ``redshifts''  for the control sample correspond to only relatively small changes in the mean color of the objects, which does not affect the calibration of the surveys greatly. We also note that the expected filter differences between the two surveys produce a relatively small and almost redshift-independent effect for the AGN sample. This is due to the small differences between the SDSS \citep[][]{Fukugita:1996} and HSC \citep[][]{Kawanomoto:2018} g-bands,\footnote{With $\lambda_{\rm eff}=4770$ \AA\ and ${\rm FWHM}=1379$ \AA\ for SDSS versus $\lambda_{\rm eff}=4754$ \AA\ and ${\rm FWHM}=1395$ \AA\ for HSC} and characteristic power-law SED of an AGN that results in modest $u-g$ and $g-r$ colors of $\approx -0.2$ to $0.3$ on the AB system \citep[e.g.,][]{Richards:2001}. Based on these two tests, we conclude that differences in the survey photometry cannot explain the observed difference between the two AGN measurements. We proceed with further tests to confirm this conclusion. \par

\subsection{Split according to brightness} \label{sec:bri}

\begin{figure*}
    \centering
    \includegraphics[width=\linewidth]{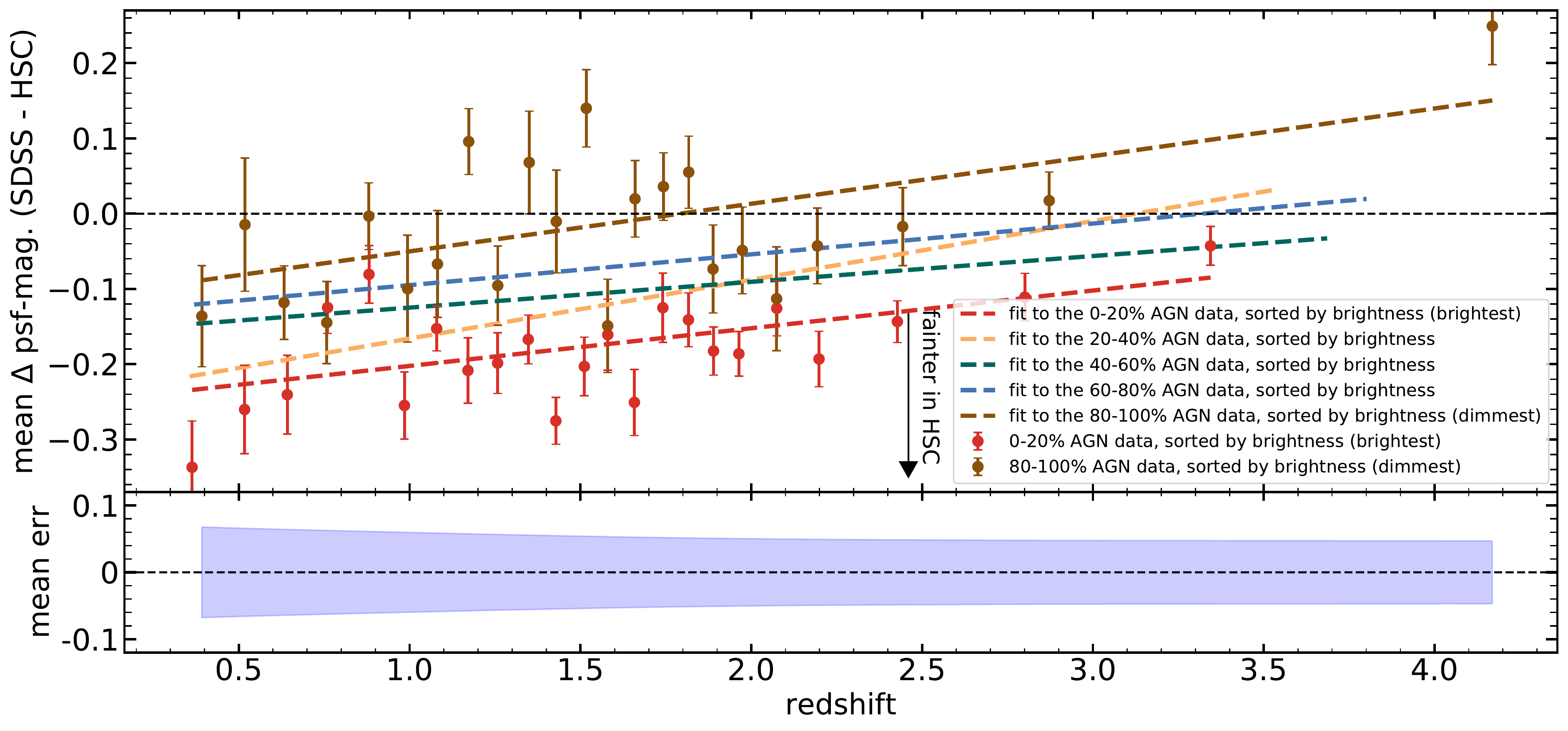}
    \caption{Mean difference in the measured psf-magnitudes in the g-band for the sample of AGN separated according to their brightness at each redshift. Different colored lines show the results of the linear fit to the subsets of the data. They have been constructed by separating the data at each redshift in quintiles, according to their brightness. Binning is coarser than in Figures \ref{fig:master} and \ref{fig:master_filter_stars} as to preserve statistical power  of  individual  data  points. We do not show the data points for the three inner quintiles of the data to improve the clarity of the figure. Lower panel shows mean 1-$\sigma$ uncertainty bands around linear fits. Note that scaling on the y-axis is different than in the upper panel.} 
    \label{fig:luminosity}
\end{figure*}

We then continue to study the redshift effect after splitting our sample in brightness. We do this for two separate reasons. Observationally, we expect that systematic differences between the surveys would be more strongly manifested for objects that have lower brightness, as various errors and uncertainties start to dominate closer to the brightness limit of the SDSS survey. Also, among less luminous AGN, which occur predominantly at lower redshifts, flux from the host galaxy may start to be nonnegligible \citep{She11}, which might bias our results. Additionally, given that variability is enhanced at lower luminosities, ``Eddington bias'', referring to the fact that intrinsically lower-luminosity AGN might get scattered into the selection of the first, shallow, survey and then ``return'' to their mean value when observed later, would produce a measured mean change of brightness that would be more noticeable at lower luminosities.  \par 
On the other hand, physically, we would expect that the mean brightness change would be larger for more luminous
AGN. Under the assumption that all AGN are the members of the same population, with the same underlying Eddington ratio distribution, AGN are bright enough to be detected in a shallow flux-limited survey only during rare parts of their life-cycle. We would expect that, on average, the population of such AGN would gravitate to their mean, low-flux state as a function of time. In particular, if this assumption is correct, we would expect that brighter AGN are in the more extreme part of their life-cycle, occupying more extreme ends of their long-term Eddington ratio distribution. We would therefore expect that brighter AGN will decrease their brightness more during any given observed time-frame. \par

We split the data in each redshift bin into five further bins, according to their observed brightness in SDSS. We then proceeded to fit the data in each of these brightness bins with a linear function and show the results of the fitting procedure in Figure \ref{fig:luminosity}. As uncertainties on the fits are quite similar for all of the 5 bins, we show the mean error on the fit in the separate panel below the main panel. We see that the effect is indeed stronger for the brighter AGN, as we expected from our theoretical reasoning. This makes us even more confident in the physical nature of this effect. We also wish to point out that the observed brightening for the dimmest objects is mostly driven by the last point at the highest redshift, and it is not obvious that it is also a physical result.

\begin{figure*}
    \centering
    \includegraphics[width=\linewidth]{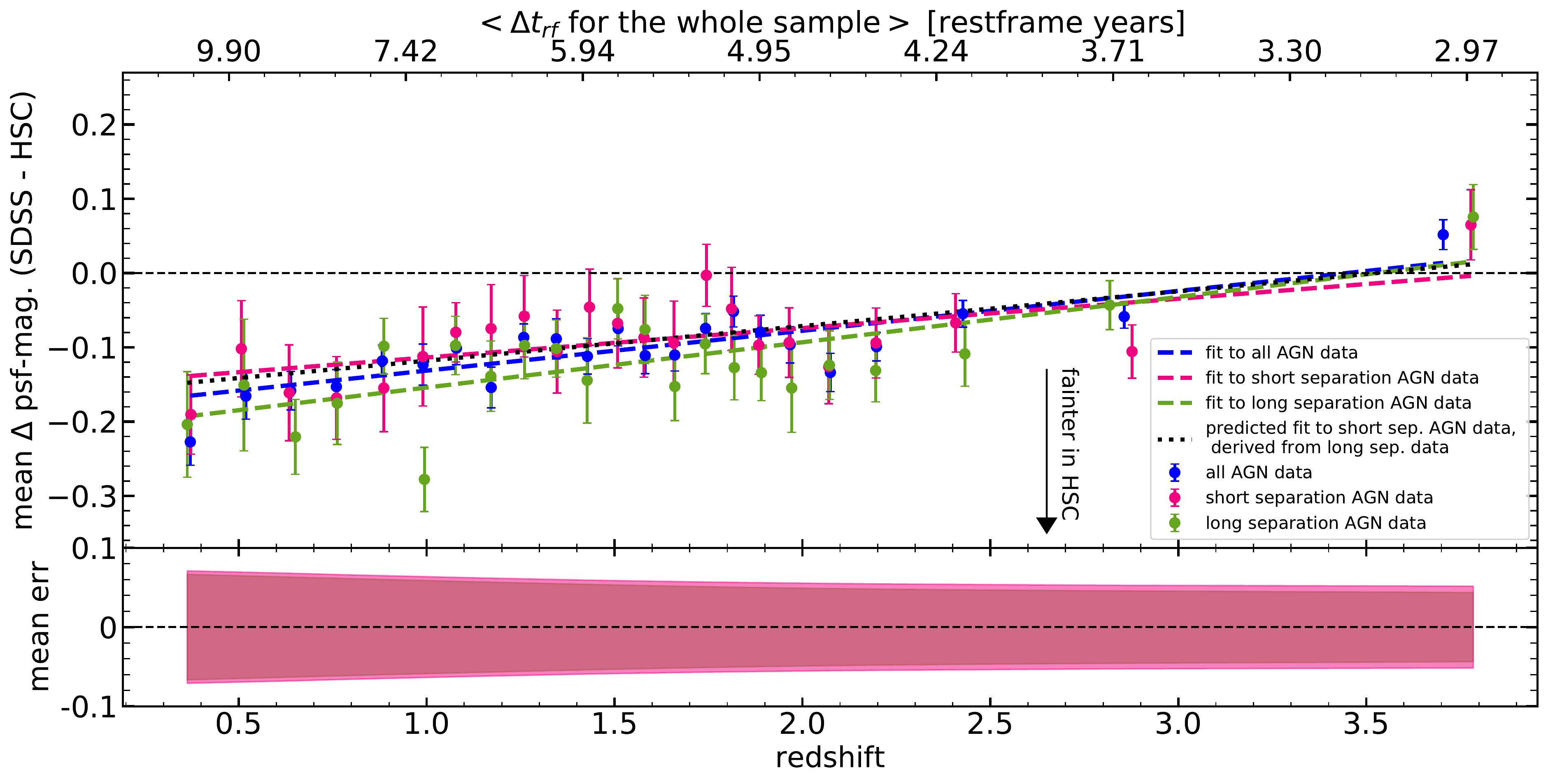}
    \caption{ Mean difference in the measured psf-magnitudes in the g-band for the sample of AGN split according to the time separation between the measurements. The blue points show the data for the whole sample of AGN in the g-band, while the blue line shows the linear fit to the data. This is equivalent to the data and fit shown in Figure \ref{fig:master}, although the binning is different to match binning for the short and long separation samples (shown in pink and green, respectively). The black dotted line shows the simplest ``derivation'' of the short separation fit, which has been calculated from the long separation fit by reducing it by the ratio of the mean time-separations for these two samples  (12.94, 16.89 years). The lower panel shows 1-$\sigma$ uncertainty bands on these linear fits for the short and long separation data.  Note that scaling on the y-axis is different than in the upper panel.   } 
    \label{fig:time}
\end{figure*}

\subsection{Split according to the time separation}\label{sec:time}

As a final check, we separated our sample in the quintiles according to the time separation between the observations. As both surveys took data over several years, we split the samples into those taken, by random chance, at the shortest and longest time intervals and compare the results. If the change of mean brightness is mostly due to observational effects, we would expect no difference between the short and long separation datasets, while if the difference is physical we would expect to see some difference between these two sets. \par
This experiment is somewhat complicated by the fact that we, at this stage, are only working with the stacked HSC data, i.e., the measured brightness of any object is a combination of measurements at different times during the duration of the survey. For HSC data, we take the mean of all of the observation times that go into each stacked observation and use that ``mean time'' as the time of the observation. The distribution of time differences between the surveys is roughly normal, with the mean at 14.85 years. We then create a sample out of the data in each redshift bin for which the time separation is within the shortest time separation quintile (short separation sample) and out of the data that are in the longest time separations quintile (long separation sample). Mean time separation for the short separation sample is 12.94 years, and for the long separation sample it is 16.89 years. \par

We then proceeded as before to study the redshift dependence of each of these samples. We show the data, the results of the linear fit to the full data and long/short separation datasets in Figure \ref{fig:time}. We see that, in general, long separation data do indeed tend to show larger changes between the two surveys. Of course, the results are quite noisy which is not surprising given the sample sizes and underlying stochastic variability. In Figure \ref{fig:time} we also show the expected linear fit for the short separation sample, which was derived from the long separation sample by multiplying the slope with the ratio of mean time separations of each sample, i.e., with the factor 12.94/16.89. This is a simplified assumption, as the mean change in brightness is not necessarily linear with time, but we see that the modifications explain well the magnitude of the observed difference. 

\section{Modeling and discussion }
\label{sec:Modelling}

\begin{figure*}
    \centering
    \includegraphics[width=\textwidth]{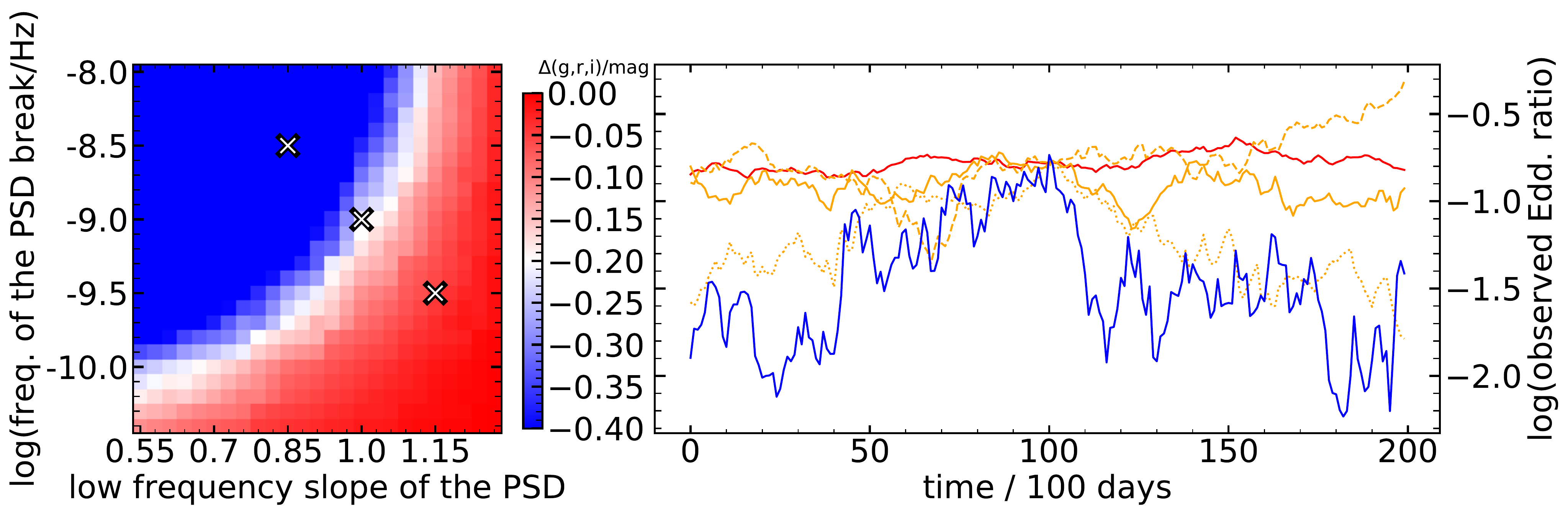}
    \caption{\textit{Left:} the mean change in measured brightness for AGN sampled at 0.5 mag above the brightness cut of a hypothetical survey, and measured again 10 rest-frame years later, as a function of $\alpha_{\rm low}$ and $f_{\rm br}$. \textit{Right:} typical simulated ``light curves'' (observed Eddington ratio) curves, from each of the areas denoted with a small cross in the left panel. Colors correspond to the colors in the left panel, where we use orange (instead of white) to color the curves from the middle, observationally plausible, region. We show three simulated curves from the observationally plausible region to demonstrate the diversity of behaviors, reminiscent of the observed diversity of AGN variability.   }
    \label{fig:delta_from_model}
\end{figure*}

In this section we discuss how the effect can be used to constrain parameters of AGN accretion given a reasonable set of assumptions. Recently \cite{Sar19} developed a code that is capable of simulating Eddington ratio curves with a duration of Myr to Gyr, and a time resolution of 10-100 days. The inputs to the code are a probability density function (PDF; in this case we assume it is the Eddington ratio function) and the power spectrum density (PSD). The assumption that PDF is given by a full Eddington ratio function, rather then by a lognormal distribution with a given $\sigma$, is the main difference from earlier modeling work (e.g., \citealp{Mac10}). We do not attempt in this work to distinguish between the two models. \par We show here an example of how the observed dependence can be used to constrain the PSD parameters. We model the PSD as a broken power, i.e., with  
\begin{equation}\label{eq:PSD}
PSD(f) = A \times \left[ \left(\frac{f}{f_{\rm br}}\right)^{\alpha_{\rm low}}  + \left(\frac{f}{f_{\rm br}}\right)^{\alpha_{\rm high}}\right]^{-1} \, 
\end{equation}

where $f_{\rm br}$ is the break frequency, and $\alpha_{\rm low}$ and $\alpha_{\rm high}$ are the slopes at lower and higher frequencies, respectively (longer and shorter timescales, respectively). While there is agreement in the community that $\alpha_{\rm high} \approx 2$ (except perhaps at shortest scales, $<$10 days, e.g., \citealp{Ede14}), the deduced values for $\alpha_{\rm low}$ and $f_{\rm br}$ vary greatly depending on the survey and method used (e.g., \citealp{Mac10,Mac12,Gra14, Koz17}). Physically, the determination of  $f_{\rm br}$ is of great interest as it would provide us with a clue about the physical scale on which the properties of AGN accretion change. \par

In the left panel of Figure \ref{fig:delta_from_model} we show the expected mean change of the measured brightness during 14.85 years, the average time difference between two measurements, as a function of $f_{\rm br}$ and $\alpha_{\rm low}$. Changing these parameters effectively changes the ``burstiness'' of the AGN accretion episodes and therefore influences how quickly the AGN are changing their luminosity in a fixed time period. This plot has been made for the systems selected with an Eddington ratio cut 0.2 dex (0.5 mag) above the break of the Eddington ratio distribution, which broadly mimics the SDSS observational cut. We can see that the observed mean brightness change defines a very specific range of allowed values in this parameter space. The two parameters are somewhat degenerate - observationally, when using a limited amount of data points, there is little difference if the process decorrelates quickly at longer timescales (small $\alpha_{\rm low}$ and large $f_{\rm br}$) or slowly at shorter timescales (large $\alpha_{\rm low}$ and small $f_{\rm br}$ - see also Figure 12 in \citealp{Cap19}). In the right-hand side of \ref{fig:delta_from_model} we show representative ``light curves'' from different regions of the parameter space. In actuality we generate curves that satisfy observed Eddington ratio distributions, and we make an assumption that variability in these ``light curves'' is equivalent to the variability in the observed light curves. In particular, we emphasize the wide variety of the behaviors for the curves that are consistent with the observed changes in the mean brightness. This is reminiscent of the wide diversity of observed variability behaviors for AGN. \par
Qualitatively, as indicated before, this model also explains why the most luminous objects at the lowest redshift are more likely to get dimmer. As they already occupy the uppermost edges of the probability density function (Eddington ratio distribution) when they were observed in SDSS, they are far more likely to get dimmer and move to more common regions of the parameter space. In other words, for the brightest AGN, the only way to go is down! \par
In the future, we aim to improve observational constraints and finely map time dependence by incorporating information from various surveys, such as POSS, Pan-STARRS, Zwicky Transient Factory and GAIA. These surveys do not achieve such depth as HSC, but monitor the sky with high cadence. We will measure the change of brightness as a function of time, while modeling the effect of the incompleteness that arises when studying AGN variability and AGN dimming in shallow surveys.
We aim to use this information describing the observed bias to distinguish between the models with different PDFs (full Eddington ratio function or lognormal distribution) and place fine constraints on the evolution of properties (primarily power spectrum density) describing AGN variability. 

\label{sec:conclusions}

\section*{Acknowledgements}
We thank the anonymous referee for providing comments. During the preparation of this manuscript, we have benefited from useful discussions with Laurent Eyer, Andy Goulding, \v Zeljko Ivezi\' c, Robert Lupton, Lauren MacArthur, Chelsea MacLeod, 
Sophie Reed, Lia Sartori, John Silvermann, and Krzysztof Suberlak. We especially thank Yusra AlSayyad, who prepared the filter flags used when retrieving the HSC data. We thank \v Zeljko Ivezi\' c and Christopher Kochanek for pointing out additional theoretical possibilities for explaining the observed effect.

This research made use of NASA's Astrophysics Data System (ADS), the arXiv.org preprint server, the Python plotting library \texttt{matplotlib} \citep{Hun07} and \texttt{astropy}, a community-developed core Python package for Astronomy \citep{Ast13}. 

\newpage
%%%%%%%%%%%%%%%%%%%%%%%%%%%%%%%%%%%%%%%%%%%%%%%%%%

%%%%%%%%%%%%%%%%%%%% REFERENCES %%%%%%%%%%%%%%%%%%

\bibliography{library}{}

\begin{thebibliography}{}
\expandafter\ifx\csname natexlab\endcsname\relax\def\natexlab#1{#1}\fi
\providecommand{\url}[1]{\href{#1}{#1}}
\providecommand{\dodoi}[1]{doi:~\href{http://doi.org/#1}{\nolinkurl{#1}}}
\providecommand{\doeprint}[1]{\href{http://ascl.net/#1}{\nolinkurl{http://ascl.net/#1}}}
\providecommand{\doarXiv}[1]{\href{https://arxiv.org/abs/#1}{\nolinkurl{https://arxiv.org/abs/#1}}}

\bibitem[{{Aihara} {et~al.}(2019){Aihara}, {AlSayyad}, {Ando}, {Armstrong},
  {Bosch}, {Egami}, {Furusawa}, {Furusawa}, {Goulding}, {Harikane}, {Hikage},
  {Ho}, {Hsieh}, {Huang}, {Ikeda}, {Imanishi}, {Ito}, {Iwata}, {Jaelani},
  {Kakuma}, {Kawana}, {Kikuta}, {Kobayashi}, {Koike}, {Komiyama}, {Li},
  {Liang}, {Lin}, {Luo}, {Lupton}, {Lust}, {MacArthur}, {Matsuoka}, {Mineo},
  {Miyatake}, {Miyazaki}, {More}, {Murata}, {Namiki}, {Nishizawa}, {Oguri},
  {Okabe}, {Okamoto}, {Okura}, {Ono}, {Onodera}, {Onoue}, {Osato}, {Ouchi},
  {Shibuya}, {Strauss}, {Sugiyama}, {Suto}, {Takada}, {Takagi}, {Takata},
  {Takita}, {Tanaka}, {Terai}, {Toba}, {Uchiyama}, {Utsumi}, {Wang}, {Wang}, \&
  {Yamada}}]{Aih19}
{Aihara}, H., {AlSayyad}, Y., {Ando}, M., {et~al.} 2019, \pasj, 106,
  \dodoi{10.1093/pasj/psz103}

\bibitem[{{Astropy Collaboration} {et~al.}(2013){Astropy Collaboration},
  {Robitaille}, {Tollerud}, {Greenfield}, {Droettboom}, {Bray}, {Aldcroft},
  {Davis}, {Ginsburg}, {Price-Whelan}, {Kerzendorf}, {Conley}, {Crighton},
  {Barbary}, {Muna}, {Ferguson}, {Grollier}, {Parikh}, {Nair}, {Unther},
  {Deil}, {Woillez}, {Conseil}, {Kramer}, {Turner}, {Singer}, {Fox}, {Weaver},
  {Zabalza}, {Edwards}, {Azalee Bostroem}, {Burke}, {Casey}, {Crawford},
  {Dencheva}, {Ely}, {Jenness}, {Labrie}, {Lim}, {Pierfederici}, {Pontzen},
  {Ptak}, {Refsdal}, {Servillat}, \& {Streicher}}]{Ast13}
{Astropy Collaboration}, {Robitaille}, T.~P., {Tollerud}, E.~J., {et~al.} 2013,
  \aap, 558, A33, \dodoi{10.1051/0004-6361/201322068}

\bibitem[{{Bauer} {et~al.}(2009){Bauer}, {Baltay}, {Coppi}, {Ellman}, {Jerke},
  {Rabinowitz}, \& {Scalzo}}]{Bau09}
{Bauer}, A., {Baltay}, C., {Coppi}, P., {et~al.} 2009, \apj, 696, 1241,
  \dodoi{10.1088/0004-637X/696/2/1241}

\bibitem[{{Bentz}(2015)}]{Ben15}
{Bentz}, M.~C. 2015, arXiv e-prints, arXiv:1505.04805.
\newblock \doarXiv{1505.04805}

\bibitem[{{Bland-Hawthorn} {et~al.}(2013){Bland-Hawthorn}, {Maloney},
  {Sutherland }, \& {Madsen}}]{Bla13}
{Bland-Hawthorn}, J., {Maloney}, P.~R., {Sutherland }, R.~S., \& {Madsen},
  G.~J. 2013, \apj, 778, 58, \dodoi{10.1088/0004-637X/778/1/58}

\bibitem[{{Caplar} {et~al.}(2017){Caplar}, {Lilly}, \& {Trakhtenbrot}}]{Cap17}
{Caplar}, N., {Lilly}, S.~J., \& {Trakhtenbrot}, B. 2017, \apj, 834, 111,
  \dodoi{10.3847/1538-4357/834/2/111}

\bibitem[{{Caplar} \& {Tacchella}(2019)}]{Cap19}
{Caplar}, N., \& {Tacchella}, S. 2019, \mnras, 487, 3845,
  \dodoi{10.1093/mnras/stz1449}

\bibitem[{{Cartier} {et~al.}(2015){Cartier}, {Lira}, {Coppi}, {S{\'a}nchez},
  {Ar{\'e}valo}, {Bauer}, {Rabinowitz}, {Zinn}, {Mu{\~n}oz}, \& {Meza}}]{Car15}
{Cartier}, R., {Lira}, P., {Coppi}, P., {et~al.} 2015, \apj, 810, 164,
  \dodoi{10.1088/0004-637X/810/2/164}

\bibitem[{{de Vries} {et~al.}(2003){de Vries}, {Becker}, \& {White}}]{deVri03}
{de Vries}, W.~H., {Becker}, R.~H., \& {White}, R.~L. 2003, \aj, 126, 1217,
  \dodoi{10.1086/377486}

\bibitem[{{de Vries} {et~al.}(2005){de Vries}, {Becker}, {White}, \&
  {Loomis}}]{deVri05}
{de Vries}, W.~H., {Becker}, R.~H., {White}, R.~L., \& {Loomis}, C. 2005, \aj,
  129, 615, \dodoi{10.1086/427393}

\bibitem[{{Edelson} {et~al.}(2014){Edelson}, {Vaughan}, {Malkan}, {Kelly},
  {Smith}, {Boyd}, \& {Mushotzky}}]{Ede14}
{Edelson}, R., {Vaughan}, S., {Malkan}, M., {et~al.} 2014, \apj, 795, 2,
  \dodoi{10.1088/0004-637X/795/1/2}

\bibitem[{{Fausnaugh} {et~al.}(2016){Fausnaugh}, {Denney}, {Barth}, {Bentz},
  {Bottorff}, {Carini}, {Croxall}, {De Rosa}, {Goad}, {Horne}, {Joner},
  {Kaspi}, {Kim}, {Klimanov}, {Kochanek}, {Leonard}, {Netzer}, {Peterson},
  {Schn{\"u}lle}, {Sergeev}, {Vestergaard}, {Zheng}, {Zu}, {Anderson},
  {Ar{\'e}valo}, {Bazhaw}, {Borman}, {Boroson}, {Brandt}, {Breeveld}, {Brewer},
  {Cackett}, {Crenshaw}, {Dalla Bont{\`a}}, {De Lorenzo-C{\'a}ceres},
  {Dietrich}, {Edelson}, {Efimova}, {Ely}, {Evans}, {Filippenko}, {Flatland},
  {Gehrels}, {Geier}, {Gelbord}, {Gonzalez}, {Gorjian}, {Grier}, {Grupe},
  {Hall}, {Hicks}, {Horenstein}, {Hutchison}, {Im}, {Jensen}, {Jones},
  {Kaastra}, {Kelly}, {Kennea}, {Kim}, {Korista}, {Kriss}, {Lee}, {Lira},
  {MacInnis}, {Manne-Nicholas}, {Mathur}, {McHardy}, {Montouri}, {Musso},
  {Nazarov}, {Norris}, {Nousek}, {Okhmat}, {Pancoast}, {Papadakis}, {Parks},
  {Pei}, {Pogge}, {Pott}, {Rafter}, {Rix}, {Saylor}, {Schimoia}, {Siegel},
  {Spencer}, {Starkey}, {Sung}, {Teems}, {Treu}, {Turner}, {Uttley},
  {Villforth}, {Weiss}, {Woo}, {Yan}, \& {Young}}]{Fau16}
{Fausnaugh}, M.~M., {Denney}, K.~D., {Barth}, A.~J., {et~al.} 2016, \apj, 821,
  56, \dodoi{10.3847/0004-637X/821/1/56}

\bibitem[{{Flesch}(2015)}]{Fle15}
{Flesch}, E.~W. 2015, \pasa, 32, e010, \dodoi{10.1017/pasa.2015.10}

\bibitem[{{Fukugita} {et~al.}(1996){Fukugita}, {Ichikawa}, {Gunn}, {Doi},
  {Shimasaku}, \& {Schneider}}]{Fukugita:1996}
{Fukugita}, M., {Ichikawa}, T., {Gunn}, J.~E., {et~al.} 1996, \aj, 111, 1748,
  \dodoi{10.1086/117915}

\bibitem[{{Graham} {et~al.}(2014){Graham}, {Djorgovski}, {Drake}, {Mahabal},
  {Chang}, {Stern}, {Donalek}, \& {Glikman}}]{Gra14}
{Graham}, M.~J., {Djorgovski}, S.~G., {Drake}, A.~J., {et~al.} 2014, \mnras,
  439, 703, \dodoi{10.1093/mnras/stt2499}

\bibitem[{{Gunn} {et~al.}(2006){Gunn}, {Siegmund}, {Mannery}, {Owen}, {Hull},
  {Leger}, {Carey}, {Knapp}, {York}, {Boroski}, {Kent}, {Lupton}, {Rockosi},
  {Evans}, {Waddell}, {Anderson}, {Annis}, {Barentine}, {Bartoszek}, {Bastian},
  {Bracker}, {Brewington}, {Briegel}, {Brinkmann}, {Brown}, {Carr},
  {Czarapata}, {Drennan}, {Dombeck}, {Federwitz}, {Gillespie}, {Gonzales},
  {Hansen}, {Harvanek}, {Hayes}, {Jordan}, {Kinney}, {Klaene}, {Kleinman},
  {Kron}, {Kresinski}, {Lee}, {Limmongkol}, {Lindenmeyer}, {Long}, {Loomis},
  {McGehee}, {Mantsch}, {Neilsen}, {Neswold}, {Newman}, {Nitta}, {Peoples},
  {Pier}, {Prieto}, {Prosapio}, {Rivetta}, {Schneider}, {Snedden}, \&
  {Wang}}]{Gun10}
{Gunn}, J.~E., {Siegmund}, W.~A., {Mannery}, E.~J., {et~al.} 2006, \aj, 131,
  2332, \dodoi{10.1086/500975}

\bibitem[{{Hunter}(2007)}]{Hun07}
{Hunter}, J.~D. 2007, Computing in Science and Engineering, 9, 90,
  \dodoi{10.1109/MCSE.2007.55}

\bibitem[{{Ivezi{\'c}} {et~al.}(2007){Ivezi{\'c}}, {Smith}, {Miknaitis}, {Lin},
  {Tucker}, {Lupton}, {Gunn}, {Knapp}, {Strauss}, {Sesar}, {Doi}, {Tanaka},
  {Fukugita}, {Holtzman}, {Kent}, {Yanny}, {Schlegel}, {Finkbeiner},
  {Padmanabhan}, {Rockosi}, {Juri{\'c}}, {Bond}, {Lee}, {Stoughton}, {Jester},
  {Harris}, {Harding}, {Morrison}, {Brinkmann}, {Schneider}, \& {York}}]{Ive07}
{Ivezi{\'c}}, {\v{Z}}., {Smith}, J.~A., {Miknaitis}, G., {et~al.} 2007, \aj,
  134, 973, \dodoi{10.1086/519976}

\bibitem[{{Johnson} {et~al.}(2018){Johnson}, {Chen}, {Straka}, {Schaye},
  {Cantalupo}, {Wendt}, {Muzahid}, {Bouch{\'e}}, {Herenz}, {Kollatschny},
  {Mulchaey}, {Marino}, {Maseda}, \& {Wisotzki}}]{Johnson:2018}
{Johnson}, S.~D., {Chen}, H.-W., {Straka}, L.~A., {et~al.} 2018, \apjl, 869,
  L1, \dodoi{10.3847/2041-8213/aaf1cf}

\bibitem[{{Kawanomoto} {et~al.}(2018){Kawanomoto}, {Uraguchi}, {Komiyama},
  {Miyazaki}, {Furusawa}, {Finet}, {Hattori}, {Wang}, {Yasuda}, \&
  {Suzuki}}]{Kawanomoto:2018}
{Kawanomoto}, S., {Uraguchi}, F., {Komiyama}, Y., {et~al.} 2018, \pasj, 70, 66,
  \dodoi{10.1093/pasj/psy056}

\bibitem[{{Koz{\l}owski}(2017)}]{Koz17}
{Koz{\l}owski}, S. 2017, \aap, 597, A128, \dodoi{10.1051/0004-6361/201629890}

\bibitem[{{Lawrence}(2018)}]{Law18}
{Lawrence}, A. 2018, Nature Astronomy, 2, 102,
  \dodoi{10.1038/s41550-017-0372-1}

\bibitem[{{MacLeod} {et~al.}(2010){MacLeod}, {Ivezi{\'c}}, {Kochanek},
  {Koz{\l}owski}, {Kelly}, {Bullock}, {Kimball}, {Sesar}, {Westman}, {Brooks},
  {Gibson}, {Becker}, \& {de Vries}}]{Mac10}
{MacLeod}, C.~L., {Ivezi{\'c}}, {\v{Z}}., {Kochanek}, C.~S., {et~al.} 2010,
  \apj, 721, 1014, \dodoi{10.1088/0004-637X/721/2/1014}

\bibitem[{{MacLeod} {et~al.}(2012){MacLeod}, {Ivezi{\'c}}, {Sesar}, {de Vries},
  {Kochanek}, {Kelly}, {Becker}, {Lupton}, {Hall}, {Richards}, {Anderson}, \&
  {Schneider}}]{Mac12}
{MacLeod}, C.~L., {Ivezi{\'c}}, {\v{Z}}., {Sesar}, B., {et~al.} 2012, \apj,
  753, 106, \dodoi{10.1088/0004-637X/753/2/106}

\bibitem[{{MacLeod} {et~al.}(2016){MacLeod}, {Ross}, {Lawrence}, {Goad},
  {Horne}, {Burgett}, {Chambers}, {Flewelling}, {Hodapp}, {Kaiser}, {Magnier},
  {Wainscoat}, \& {Waters}}]{Mac16}
{MacLeod}, C.~L., {Ross}, N.~P., {Lawrence}, A., {et~al.} 2016, \mnras, 457,
  389, \dodoi{10.1093/mnras/stv2997}

\bibitem[{{Miyazaki} {et~al.}(2018){Miyazaki}, {Komiyama}, {Kawanomoto}, {Doi},
  {Furusawa}, {Hamana}, {Hayashi}, {Ikeda}, {Kamata}, {Karoji}, {Koike},
  {Kurakami}, {Miyama}, {Morokuma}, {Nakata}, {Namikawa}, {Nakaya}, {Nariai},
  {Obuchi}, {Oishi}, {Okada}, {Okura}, {Tait}, {Takata}, {Tanaka}, {Tanaka},
  {Terai}, {Tomono}, {Uraguchi}, {Usuda}, {Utsumi}, {Yamada}, {Yamanoi},
  {Aihara}, {Fujimori}, {Mineo}, {Miyatake}, {Oguri}, {Uchida}, {Tanaka},
  {Yasuda}, {Takada}, {Murayama}, {Nishizawa}, {Sugiyama}, {Chiba}, {Futamase},
  {Wang}, {Chen}, {Ho}, {Liaw}, {Chiu}, {Ho}, {Lai}, {Lee}, {Jeng}, {Iwamura},
  {Armstrong}, {Bickerton}, {Bosch}, {Gunn}, {Lupton}, {Loomis}, {Price},
  {Smith}, {Strauss}, {Turner}, {Suzuki}, {Miyazaki}, {Muramatsu}, {Yamamoto},
  {Endo}, {Ezaki}, {Ito}, {Kawaguchi}, {Sofuku}, {Taniike}, {Akutsu}, {Dojo},
  {Kasumi}, {Matsuda}, {Imoto}, {Miwa}, {Suzuki}, {Takeshi}, \&
  {Yokota}}]{Miy18}
{Miyazaki}, S., {Komiyama}, Y., {Kawanomoto}, S., {et~al.} 2018, \pasj, 70, S1,
  \dodoi{10.1093/pasj/psx063}

\bibitem[{{Morganson} {et~al.}(2014){Morganson}, {Burgett}, {Chambers},
  {Green}, {Kaiser}, {Magnier}, {Marshall}, {Morgan}, {Price}, {Rix},
  {Schlafly}, {Tonry}, \& {Walter}}]{Mor14}
{Morganson}, E., {Burgett}, W.~S., {Chambers}, K.~C., {et~al.} 2014, \apj, 784,
  92, \dodoi{10.1088/0004-637X/784/2/92}

\bibitem[{{Novak} {et~al.}(2011){Novak}, {Ostriker}, \& {Ciotti}}]{Nov11}
{Novak}, G.~S., {Ostriker}, J.~P., \& {Ciotti}, L. 2011, \apj, 737, 26,
  \dodoi{10.1088/0004-637X/737/1/26}

\bibitem[{{Richards} {et~al.}(2001){Richards}, {Fan}, {Schneider}, {Vanden
  Berk}, {Strauss}, {York}, {Anderson}, {Anderson}, {Annis}, {Bahcall},
  {Bernardi}, {Briggs}, {Brinkmann}, {Brunner}, {Burles}, {Carey}, {Castand
  er}, {Connolly}, {Crocker}, {Csabai}, {Doi}, {Finkbeiner}, {Friedman},
  {Frieman}, {Fukugita}, {Gunn}, {Hindsley}, {Ivezi{\'c}}, {Kent}, {Knapp},
  {Lamb}, {Leger}, {Long}, {Loveday}, {Lupton}, {McKay}, {Meiksin}, {Merrelli},
  {Munn}, {Newberg}, {Newcomb}, {Nichol}, {Owen}, {Pier}, {Pope}, {Richmond},
  {Rockosi}, {Schlegel}, {Siegmund}, {Smee}, {Snir}, {Stoughton}, {Stubbs},
  {SubbaRao}, {Szalay}, {Szokoly}, {Tremonti}, {Uomoto}, {Waddell}, {Yanny}, \&
  {Zheng}}]{Richards:2001}
{Richards}, G.~T., {Fan}, X., {Schneider}, D.~P., {et~al.} 2001, \aj, 121,
  2308, \dodoi{10.1086/320392}

\bibitem[{{Rumbaugh} {et~al.}(2018){Rumbaugh}, {Shen}, {Morganson}, {Liu},
  {Banerji}, {McMahon}, {Abdalla}, {Benoit-L{\'e}vy}, {Bertin}, {Brooks},
  {Buckley-Geer}, {Capozzi}, {Carnero Rosell}, {Carrasco Kind}, {Carretero},
  {Cunha}, {D'Andrea}, {da Costa}, {DePoy}, {Desai}, {Doel}, {Frieman},
  {Garc{\'\i}a-Bellido}, {Gruen}, {Gruendl}, {Gschwend}, {Gutierrez},
  {Honscheid}, {James}, {Kuehn}, {Kuhlmann}, {Kuropatkin}, {Lima}, {Maia},
  {Marshall}, {Martini}, {Menanteau}, {Plazas}, {Reil}, {Roodman}, {Sanchez},
  {Scarpine}, {Schindler}, {Schubnell}, {Sheldon}, {Smith}, {Soares-Santos},
  {Sobreira}, {Suchyta}, {Swanson}, {Walker}, {Wester}, \& {(DES
  Collaboration}}]{Rum18}
{Rumbaugh}, N., {Shen}, Y., {Morganson}, E., {et~al.} 2018, \apj, 854, 160,
  \dodoi{10.3847/1538-4357/aaa9b6}

\bibitem[{{Sartori} {et~al.}(2018){Sartori}, {Schawinski}, {Trakhtenbrot},
  {Caplar}, {Treister}, {Koss}, {Urry}, \& {Zhang}}]{Sar18}
{Sartori}, L.~F., {Schawinski}, K., {Trakhtenbrot}, B., {et~al.} 2018, \mnras,
  476, L34, \dodoi{10.1093/mnrasl/sly025}

\bibitem[{{Sartori} {et~al.}(2019){Sartori}, {Trakhtenbrot}, {Schawinski},
  {Caplar}, {Treister}, \& {Zhang}}]{Sar19}
{Sartori}, L.~F., {Trakhtenbrot}, B., {Schawinski}, K., {et~al.} 2019, \apj,
  883, 139, \dodoi{10.3847/1538-4357/ab3c55}

\bibitem[{{Sartori} {et~al.}(2016){Sartori}, {Schawinski}, {Koss}, {Treister},
  {Maksym}, {Keel}, {Urry}, {Lintott}, \& {Wong}}]{Sar16}
{Sartori}, L.~F., {Schawinski}, K., {Koss}, M., {et~al.} 2016, \mnras, 457,
  3629, \dodoi{10.1093/mnras/stw230}

\bibitem[{{Schmidt} {et~al.}(2010){Schmidt}, {Marshall}, {Rix}, {Jester},
  {Hennawi}, \& {Dobler}}]{Schmi10}
{Schmidt}, K.~B., {Marshall}, P.~J., {Rix}, H.-W., {et~al.} 2010, \apj, 714,
  1194, \dodoi{10.1088/0004-637X/714/2/1194}

\bibitem[{{Schmidt} {et~al.}(2018){Schmidt}, {Hennawi}, {Worseck}, {Davies},
  {Luki{\'c}}, \& {O{\~n}orbe}}]{Schmidt:2018}
{Schmidt}, T.~M., {Hennawi}, J.~F., {Worseck}, G., {et~al.} 2018, \apj, 861,
  122, \dodoi{10.3847/1538-4357/aac8e4}

\bibitem[{{Schneider} {et~al.}(2010){Schneider}, {Richards}, {Hall}, {Strauss},
  {Anderson}, {Boroson}, {Ross}, {Shen}, {Brandt}, {Fan}, {Inada}, {Jester},
  {Knapp}, {Krawczyk}, {Thakar}, {Vanden Berk}, {Voges}, {Yanny}, {York},
  {Bahcall}, {Bizyaev}, {Blanton}, {Brewington}, {Brinkmann}, {Eisenstein},
  {Frieman}, {Fukugita}, {Gray}, {Gunn}, {Hibon}, {Ivezi{\'c}}, {Kent}, {Kron},
  {Lee}, {Lupton}, {Malanushenko}, {Malanushenko}, {Oravetz}, {Pan}, {Pier},
  {Price}, {Saxe}, {Schlegel}, {Simmons}, {Snedden}, {SubbaRao}, {Szalay}, \&
  {Weinberg}}]{Schne10}
{Schneider}, D.~P., {Richards}, G.~T., {Hall}, P.~B., {et~al.} 2010, \aj, 139,
  2360, \dodoi{10.1088/0004-6256/139/6/2360}

\bibitem[{{Shen} {et~al.}(2011){Shen}, {Richards}, {Strauss}, {Hall},
  {Schneider}, {Snedden}, {Bizyaev}, {Brewington}, {Malanushenko},
  {Malanushenko}, {Oravetz}, {Pan}, \& {Simmons}}]{She11}
{Shen}, Y., {Richards}, G.~T., {Strauss}, M.~A., {et~al.} 2011, \apjs, 194, 45,
  \dodoi{10.1088/0067-0049/194/2/45}

\bibitem[{{Smith} {et~al.}(2018){Smith}, {Mushotzky}, {Boyd}, {Malkan},
  {Howell}, \& {Gelino}}]{Smi18}
{Smith}, K.~L., {Mushotzky}, R.~F., {Boyd}, P.~T., {et~al.} 2018, \apj, 857,
  141, \dodoi{10.3847/1538-4357/aab88d}

\bibitem[{{Vanden Berk} {et~al.}(2001){Vanden Berk}, {Richards}, {Bauer},
  {Strauss}, {Schneider}, {Heckman}, {York}, {Hall}, {Fan}, {Knapp},
  {Anderson}, {Annis}, {Bahcall}, {Bernardi}, {Briggs}, {Brinkmann}, {Brunner},
  {Burles}, {Carey}, {Castander}, {Connolly}, {Crocker}, {Csabai}, {Doi},
  {Finkbeiner}, {Friedman}, {Frieman}, {Fukugita}, {Gunn}, {Hennessy},
  {Ivezi{\'c}}, {Kent}, {Kunszt}, {Lamb}, {Leger}, {Long}, {Loveday}, {Lupton},
  {Meiksin}, {Merelli}, {Munn}, {Newberg}, {Newcomb}, {Nichol}, {Owen}, {Pier},
  {Pope}, {Rockosi}, {Schlegel}, {Siegmund}, {Smee}, {Snir}, {Stoughton},
  {Stubbs}, {SubbaRao}, {Szalay}, {Szokoly}, {Tremonti}, {Uomoto}, {Waddell},
  {Yanny}, \& {Zheng}}]{Vanden-Berk:2001}
{Vanden Berk}, D.~E., {Richards}, G.~T., {Bauer}, A., {et~al.} 2001, \aj, 122,
  549, \dodoi{10.1086/321167}

\bibitem[{{Voevodkin}(2011)}]{Voe11}
{Voevodkin}, A. 2011, arXiv e-prints, arXiv:1107.4244.
\newblock \doarXiv{1107.4244}

\bibitem[{{York} {et~al.}(2000){York}, {Adelman}, {Anderson}, {Anderson},
  {Annis}, {Bahcall}, {Bakken}, {Barkhouser}, {Bastian}, {Berman}, {Boroski},
  {Bracker}, {Briegel}, {Briggs}, {Brinkmann}, {Brunner}, {Burles}, {Carey},
  {Carr}, {Castander}, {Chen}, {Colestock}, {Connolly}, {Crocker}, {Csabai},
  {Czarapata}, {Davis}, {Doi}, {Dombeck}, {Eisenstein}, {Ellman}, {Elms},
  {Evans}, {Fan}, {Federwitz}, {Fiscelli}, {Friedman}, {Frieman}, {Fukugita},
  {Gillespie}, {Gunn}, {Gurbani}, {de Haas}, {Haldeman}, {Harris}, {Hayes},
  {Heckman}, {Hennessy}, {Hindsley}, {Holm}, {Holmgren}, {Huang}, {Hull},
  {Husby}, {Ichikawa}, {Ichikawa}, {Ivezi{\'c}}, {Kent}, {Kim}, {Kinney},
  {Klaene}, {Kleinman}, {Kleinman}, {Knapp}, {Korienek}, {Kron}, {Kunszt},
  {Lamb}, {Lee}, {Leger}, {Limmongkol}, {Lindenmeyer}, {Long}, {Loomis},
  {Loveday}, {Lucinio}, {Lupton}, {MacKinnon}, {Mannery}, {Mantsch}, {Margon},
  {McGehee}, {McKay}, {Meiksin}, {Merelli}, {Monet}, {Munn}, {Narayanan},
  {Nash}, {Neilsen}, {Neswold}, {Newberg}, {Nichol}, {Nicinski}, {Nonino},
  {Okada}, {Okamura}, {Ostriker}, {Owen}, {Pauls}, {Peoples}, {Peterson},
  {Petravick}, {Pier}, {Pope}, {Pordes}, {Prosapio}, {Rechenmacher}, {Quinn},
  {Richards}, {Richmond}, {Rivetta}, {Rockosi}, {Ruthmansdorfer}, {Sand ford},
  {Schlegel}, {Schneider}, {Sekiguchi}, {Sergey}, {Shimasaku}, {Siegmund},
  {Smee}, {Smith}, {Snedden}, {Stone}, {Stoughton}, {Strauss}, {Stubbs},
  {SubbaRao}, {Szalay}, {Szapudi}, {Szokoly}, {Thakar}, {Tremonti}, {Tucker},
  {Uomoto}, {Vanden Berk}, {Vogeley}, {Waddell}, {Wang}, {Watanabe},
  {Weinberg}, {Yanny}, {Yasuda}, \& {SDSS Collaboration}}]{York:2000}
{York}, D.~G., {Adelman}, J., {Anderson}, John~E., J., {et~al.} 2000, \aj, 120,
  1579, \dodoi{10.1086/301513}

\end{thebibliography}
\bibliographystyle{aasjournal}

%%%%%%%%%%%%%%%%% APPENDICES %%%%%%%%%%%%%%%%%%%%%

\label{lastpage}
\end{document}